\documentclass[11pt,preprintnumbers,nofootinbib]{revtex4-2}

\usepackage{soul}
\usepackage{multirow}
\usepackage{booktabs} 

\usepackage[colorlinks=true, pdfstartview=FitV, linkcolor=blue, citecolor=red, urlcolor=magenta]{hyperref}

\usepackage{graphicx}
\usepackage{latexsym}
\usepackage{amsmath}
\usepackage{amsfonts}
\usepackage{amssymb}

\usepackage{verbatim}
\usepackage{dcolumn}
\usepackage{amssymb}
\usepackage{bm}
\usepackage{float}
\usepackage{physics}
\usepackage[english]{babel}
\usepackage{subfigure}
\usepackage{todonotes}
\usepackage{footnote}
\usepackage{hyperref}

\begin{document}

\newcommand{\NITK}{
\affiliation{Department of Physics, National Institute of Technology Karnataka, Surathkal  575 025, India}
}

\newcommand{\IIT}{\affiliation{Department of Physics, Indian Institute of Technology, Ropar, Rupnagar, Punjab 140 001, India}}

\newcommand{\WB}{\affiliation{Department of Physics, Mahishadal Raj College, West Bengal, 721628 India}}

\newcommand{\KL}{\affiliation{Department of Physics, Kannur University, Payyanur, Kerala, 6710327, Kerala, India}}

\newcommand{\Shreyasusa}{\affiliation {Department of Oral Health Sciences, School of Dentistry,\\
University of Washington, Seattle, WA 98195, USA.}}

\title{Thermodynamics, photon sphere and thermodynamic geometry of Ay\'{o}n-Beato-Garc\'{i}a Spacetime}

\author{Kartheek Hegde}
\email{hegde.kartheek@gmail.com}
\NITK

\author{A. Naveena Kumara}
\email{naviphysics@gmail.com}
\NITK

\author{C. L. Ahmed Rizwan}
\email{ahmedrizwancl@gmail.com}
\NITK
\KL

\author{Md Sabir Ali}
\email{alimd.sabir3@gmail.com}
\IIT
\WB

\author{Shreyas Punacha}
\email{shreyasp444@gmail.com, shreyas4@uw.edu}
\Shreyasusa

\author{K. M. Ajith}
\email{ajith@nitk.edu.in}
\NITK

\begin{abstract}
We study the thermodynamics of the Ay\'{o}n-Beato-Garc\'{i}a black hole and the relationship between photon orbits and the thermodynamic phase transitions of the black hole in AdS spacetime. We then examine the interactions between the microstructures of the black hole using Ruppeiner geometry. The radius of the photon orbit and the minimum impact parameter behave non-monotonically below the critical point, mimicking the behaviour of Hawking temperature and pressure in extended thermodynamics. Their changes during the large black hole--small black hole phase transition serve as the order parameter, possessing a critical exponent of $1/2$. The results demonstrate that the gravity and thermodynamics of the Ay\'{o}n-Beato-Garc\'{i}a black hole are closely related. Furthermore, we explore the thermodynamic geometry, which provides insight into the microstructure interactions of the black hole. We find that the large black hole phase is analogous to a bosonic gas with a dominant attractive interaction, while the small black hole phase behaves like an anyonic gas with both attractive and repulsive interactions.
\end{abstract}


\maketitle

\section{Introduction}
Black holes are now recognized not only as purely gravitational objects but also as thermodynamic systems, with well-defined temperature and entropy identified from their surface gravity and horizon area, respectively \cite{Bekenstein1973, Bardeen:1973gs}. This recognition has given rise to the notion of black hole thermodynamics, wherein the laws of black hole mechanics mirror those of ordinary thermodynamics~\cite{Bardeen1973, Witten:2024upt}. One of the most intriguing developments in this field is the concept of black hole chemistry: by treating the cosmological constant as a thermodynamic pressure term, a more complete “extended thermodynamics” emerges \cite{Kastor:2009wy, Dolan:2011xt}, giving rise to a rich phase structure reminiscent of that observed in standard thermodynamic systems such as the van der Waals fluid \cite{Kubiznak2012, Gunasekaran2012, Kubiznak:2016qmn}.

A valuable approach to understanding the characteristic features of a gravitational system is to examine how test particles, especially photons, behave in its vicinity. Photons approaching a black hole can under suitable conditions orbit around it, forming a photon sphere that has important theoretical and observational implications for black hole shadows, gravitational lensing, and quasinormal modes (QNMs) \cite{Wu:2021pgf,Liu:2023kxd,Lin:2024ubg}. Wei \emph{et al.} \cite{Wei:2017mwc, Wei:2018aqm} made the significant discovery that parameters related to these photon orbits encode information about the black hole’s phase transition, at least in asymptotically anti-de Sitter (AdS) spacetimes. Specifically, the photon orbit radius and minimum impact parameter exhibit discontinuities or sudden changes at phase transition points, suggesting they can serve as order parameters. This intriguing connection has been explored in various contexts, including Kerr-AdS black holes \cite{Wei:2018aqm}, Born–Infeld-AdS black holes \cite{Xu:2019yub}, regular AdS black holes \cite{A.:2019mqv}, black holes in massive gravity \cite{Chabab:2019kfs}, Born–Infeld–dilaton black holes \cite{Li:2019dai}, and five-dimensional Gauss–Bonnet black holes \cite{Han:2018ooi}, among others \cite{Zhang:2019tzi, Bhamidipati:2018yqy, Wei:2019jve}.

Another notable development in black hole thermodynamics is the effort to probe the system’s microstructure. A powerful tool in this context is the Ruppeiner geometry~\cite{Ruppeinerb2008}, which constructs a thermodynamic line element in the parameter space of fluctuating coordinates (often chosen as temperature and volume) \cite{Wei2019a, Wei2019b, HosseiniMansoori:2019jcs}. The resulting curvature scalar can reveal the nature (attractive or repulsive) and strength of microscopic interactions, as indicated by its sign and magnitude \cite{Ruppeiner95, Janyszek_1990, Oshima_1999x, Mirza2008, PhysRevE.88.032123}. Although various black hole systems exhibit phase behaviours analogous to the van der Waals fluid, their microscopic interactions can differ significantly. For instance, charged AdS black holes show mixed repulsive-attractive interactions \cite{Wei2019a, Wei2019b}, whereas certain Gauss–Bonnet black holes display purely attractive interactions \cite{Wei:2019ctz}. Many subsequent studies have applied Ruppeiner geometry to reveal how microscopic interactions underlie the observed phase structure of diverse black hole families \cite{Wei2015, Guo2019, Miao2017, Zangeneh2017, Kumara:2019xgt, Kumara:2020mvo, Kumara:2020ucr, Dehyadegari:2020ebz, NaveenaKumara:2020hov, NaveenaKumara:2020biu, Rizwan:2020bhp, Xu:2019nnp, Chabab2018, Deng2017, Miao2019a, Chen2019, Du2019, Dehyadegari2017, Ghosh:2019pwy, Ghosh:2020kba}.

A particularly compelling direction in black hole physics involves regular (or singularity-free) solutions to Einstein’s equations. Traditionally, black holes contain a physical singularity at their core, which is often viewed as a breakdown of classical gravity. In 1968, Bardeen proposed a static, spherically symmetric black hole model without a central singularity \cite{1968qtr..conf...87B}, although at first it was not recognized as an exact solution to Einstein’s equations. Later, Ayón-Beato and García made crucial progress by deriving exact, singularity-free solutions via nonlinear electrodynamics (NLED) coupled to Einstein gravity \cite{Ayon-Beato:1998hmi, Ayon-Beato:1999qin, Ayon-Beato:1999kuh}. These solutions feature a proper Maxwell limit in the weak-field regime and satisfy the weak energy condition. They further demonstrated that Bardeen’s original black hole could be interpreted as a nonlinear magnetic monopole solution, thereby making Bardeen’s model an exact NLED-based solution \cite{Ayon-Beato:2000mjt} \footnote{There exist numerous other regular black hole solutions \cite{Dymnikova:1992ux, Dymnikova1996, Mars:1996khm, Borde:1996df, Mbonye:2005im, Hayward:2005gi}.}.

Motivated by the quest for singularity-free models, we focus here on the AdS counterpart of the Ayón-Beato–García (ABG) black hole and undertake a detailed examination of its thermodynamics, photon orbit properties, and microscopic interactions. Although the ABG-AdS black hole is a fully regular solution, a comprehensive thermodynamic investigation in extended phase space has been largely absent from the literature until recently \cite{Singh:2021nvm}. We extend the study by analysing its phase structure, identifying possible phase transitions through the behaviour of photon orbit parameters, and exploring the microstructure using Ruppeiner geometry. Recently there have been many studies on gravity coupled to NLED solutions, such as \cite{Kumar:2024cnh,Belhaj:2022gcj,Rehan:2024dsg,Habibina:2020msd,Allahyari:2019jqz,Ramadhan:2023ogm,Du:2022quq}.

This article is organized as follows. In Section \ref{sec2}, we introduce the ABG-AdS black hole and analyse its thermodynamics, including its phase structure. In Section \ref{sec3}, we investigate the photon orbits and demonstrate how their characteristic parameters reflect the black hole’s phase transitions. Section \ref{sec4} presents the Ruppeiner geometric analysis, revealing insights into the microscopic interactions that drive these phase transitions. We summarize our key findings and offer concluding remarks in Section \ref{sec5}.


\section{Phase structure of the Ay\'{o}n Beato-Garc\'{i}a AdS Black Hole}
\label{sec2}

We begin by introducing the Ay\'{o}n Beato-Garc\'{i}a (ABG) black hole spacetime. This spacetime arises within the framework of standard General Relativity, where singularity-free solutions to the Einstein field equations can be achieved by coupling to an appropriate form of nonlinear electrodynamics. In the weak-field limit, this nonlinear electrodynamics reduces to the conventional linear Maxwell theory, as discussed in \cite{Toshmatov:2014nya, Ayon-Beato:1998hmi} \footnote{For discussions and resolutions of no-go theorem see Refs. \cite{Bronnikov:2000yz,Burinskii:2002pz}.}. 
The action for Einstein gravity minimally coupled to a NLED source in a spacetime with cosmological constant $\Lambda$ is given by, 
\begin{equation}
\label{action}
    \mathcal{S}= \frac{1}{16 \pi} \int{{d^4 x}\sqrt{-g} \left[R - 2\Lambda - 4 \mathcal{L\left(F\right)} \right]},
\end{equation}
where $R$ and ${g}$ are the Ricci scalar and the determinant of the metric tensor, respectively. $\mathcal{L(F)}$ is the Lagrangian density of nonlinear electrodynamics which is a function of $\mathcal{F} = (1/4) F_{\mu \nu }F^{\mu \nu }$, given by,
\begin{equation}
\mathcal{L\left(F\right)} = \frac{\mathcal{F} (1 - 3 \sqrt{2 {q^2} \mathcal{F}})}{(1 + \sqrt{2 {q^2} \mathcal{F}})^3} - \frac{3 M}{q^3} \left[\frac{(2 {q^2} \mathcal{F})^{5/4}}{(1 + \sqrt{2 {q^2} \mathcal{F}})^{5/2}} \right] ,
\end{equation}
with $F_{\mu\nu}=2 \delta{^\theta}{_{[\mu}} \delta{^\phi}{_{\nu]}} \chi(r,\theta)$, for a purely magnetically charged black hole and for this pure magnetic charge only $F_{\theta \phi}$ survives,
\begin{equation}
F_{\theta \phi}= q(r) \sin{\theta}, \qquad A{_\phi} = - q(r) \cos{\theta}.
\end{equation}
Using, $d \mathcal{F} = 0$, we get,
\begin{equation}
F_{\theta \phi}= q \sin{\theta}.
\end{equation}
The Maxwell invariant is,
\begin{equation}
\mathcal{F} = \frac{q^2}{2 r^4}
\end{equation}
The resulting metric is,
\begin{equation}
ds^2 = -f(r)\,dt^2 + \frac{dr^2}{f(r)} + r^2\,d\Omega^2,
\end{equation}
where $d\Omega^2 = d\theta^2 + \sin^2 \theta\, d\varphi^2$ is the 2-sphere, and the metric function is
\begin{equation}\label{metric function}
f(r) = 1 - \frac{2 M r^2}{(q^2 + r^2)^{3/2}} + \frac{q^2 r^2}{(q^2 + r^2)^2} - \frac{\Lambda r^2}{3}.
\end{equation}
    
The spacetime is regular, with no singularity at the centre of the black hole when $r = 0$. This can be verified by observing that the Kretschmann scalar does not have a singularity at $r = 0$,
\begin{equation}
k = R^{\mu \nu \rho \sigma} R_{\mu \nu \rho \sigma}
= \frac{8 \Big(36 M^2 + 12 M q (\Lambda q^2 - 3) + q^2 (\Lambda q^2 - 3)^2 \Big)}{3 q^6}.
\end{equation}

We note that some aspects of thermodynamics were studied in Ref.~\cite{Singh:2021nvm}, where the authors investigated the phase transition of ABG black holes using the behaviour of specific heat and calculated the critical exponents. We provide a comprehensive treatment here with additional inputs. We present the coexistence and spinodal curves, which are essential for analysing the photon orbit. We also show that the horizon radius can be treated as an order parameter. Additionally, we have calculated the thermodynamic quantities and studied the phase transition properties and phase structure of the ABG black hole for completeness. This analysis is also necessary for the study of thermodynamic geometry and the correlation between gravity and thermodynamics through the study of the photon sphere around the ABG black hole.

We study the extended thermodynamics by considering the dynamic cosmological constant as the thermodynamic pressure, $P = -\Lambda / 8\pi$. We obtain the mass of the black hole from (\ref{metric function}),
\begin{equation}
M = \frac{q^4 \bigl(3 - \Lambda r_h^2\bigr) + q^2 \bigl(9 r_h^2 - 2 \Lambda r_h^4\bigr)
- \Lambda r_h^6 + 3 r_h^4}{6\,r_h^2\,\sqrt{q^2 + r_h^2}},
\end{equation}
where $r_h$ is the radius of the event horizon of the black hole.
The Hawking temperature is obtained from the surface gravity $\kappa$ as, $T = \kappa / 2\pi$,
\begin{align}
T = \frac{1}{4 \pi \, r_h \,\bigl(q^2 + r_h^2\bigr)^3} \Bigg[q^2 r_h^4 \bigl(16 \pi P\,r_h^2 - 1\bigr)
- 2 q^6 + r_h^6 + 8 \pi P\,r_h^8 + q^4 r_h^2 \bigl(8 \pi P\,r_h^2 - 3\bigr)\Bigg].
\label{temperature}
\end{align}
The volume of the black hole is calculated from the first law,
\begin{equation}
V =\left( \frac{\partial M}{\partial P} \right)_{S,Q}= \frac{4}{3}\,\pi\,\bigl(q^2 + r_h^2\bigr)^{3/2},
\end{equation}

and the entropy, as obtained by the first law of thermodynamics, is,
\begin{align}
S = \int \frac{dM}{T} = 2\pi \Bigg[\Bigl(\frac{r}{2} - \frac{q^2}{r}\Bigr)\sqrt{q^2 + r^2}
+ \frac{3}{2} q^2 \log\bigl(\sqrt{q^2 + r^2} + r\bigr)\Bigg].
\end{align}
These quantities satisfy the first law of black hole mechanics,
\begin{equation}
dM = T\,dS + \Psi\,dQ + V\,dP,
\end{equation}
and its integral form, the Smarr relation,
\begin{equation}
M = 2(TS - VP) + \Psi Q.
\end{equation}
The entropy $S$ follows from the first law and not from the Bekenstein's area law, which is characteristic of some regular black holes~\citep{Ma:2014qma, Nam:2018sii}. This does not affect the phase transition, microstructure, or other related properties of the black hole; the qualitative analysis remains unchanged.

The black hole solution we consider is spherically symmetric, where entropy and volume are interdependent, and hence the heat capacity vanishes,
\begin{equation}
C_V = T \left( \frac{\partial S}{\partial T}\right)_V = 0.
\label{cv}
\end{equation}
The equation of state is obtained by rearranging the expression for Hawking temperature (Eq.~\ref{temperature}),
\begin{align}
P = \frac{1}{8 \pi\,r_h^4\,\bigl(q^2 + r_h^2\bigr)^2} &\Bigg[
3 q^4 r_h^2 (4 \pi r_h T + 1) + q^2 r_h^4 (12 \pi r_h T + 1)
\nonumber\\
&+ q^6 (4 \pi r_h T + 2) + r_h^6 (4 \pi r_h T - 1)
\Bigg].
\end{align}

\begin{figure*}[t]
\centering
\includegraphics[width=\textwidth]{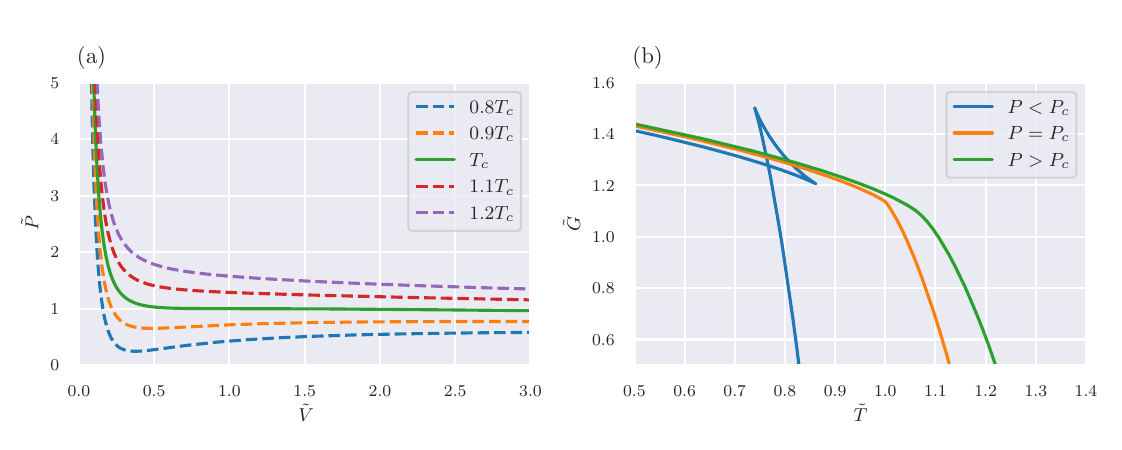}
\caption{(a) Reduced pressure ($\tilde{P}$) versus reduced volume ($\tilde{V}$) isotherms for different temperatures. The isotherms display oscillatory behaviour for temperatures below the critical temperature ($T_c$), which disappears for temperatures above $T_c$. (b) Reduced Gibbs free energy ($\tilde{G}$) versus reduced temperature ($\tilde{T}$) plots for various pressures ($P$). A swallowtail behaviour is observed for pressures below the critical pressure ($P_c$), characteristic of a van der Waals-like phase transition.}
\label{fig1}
\end{figure*}

\begin{figure*}[t]
\centering
\includegraphics[width=\textwidth]{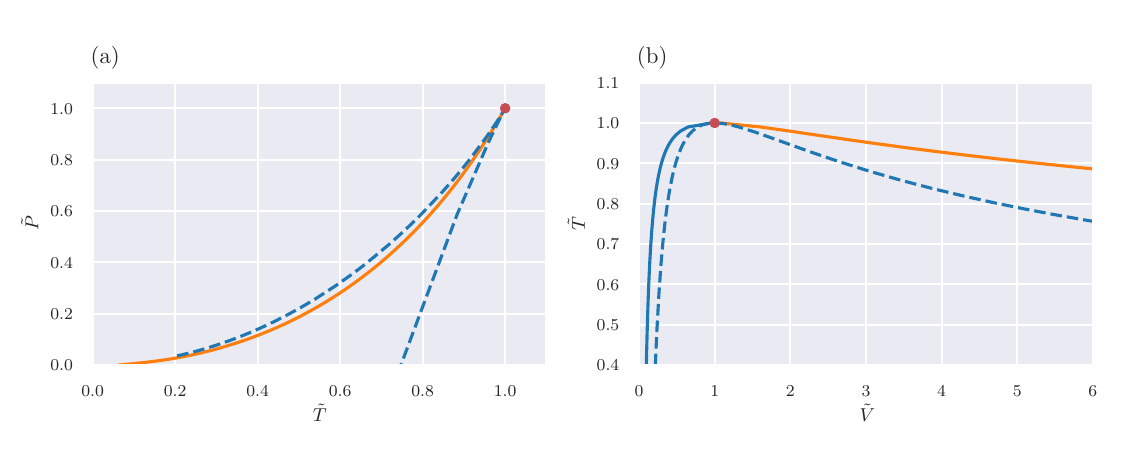}
\caption{Phase structure of the ABG AdS black hole. (a) In the $P$-$T$ plane, the solid orange curve represents the coexistence curve, while the dashed blue curves depict the spinodal curves. The region above the coexistence curve corresponds to the stable large black hole (LBH) phase, and below it to the stable small black hole (SBH) phase. The area between the spinodal curves indicates metastable phases. (b) In the $T$-$V$ plane, the area above the coexistence curve signifies unstable phases, and the area below denotes stable phases. The region between the coexistence and spinodal curves represents metastable phases. The critical temperature is marked by a red dot.}
\label{fig2}
\end{figure*}

\begin{figure*}[t]
\centering
\includegraphics[width=\textwidth]{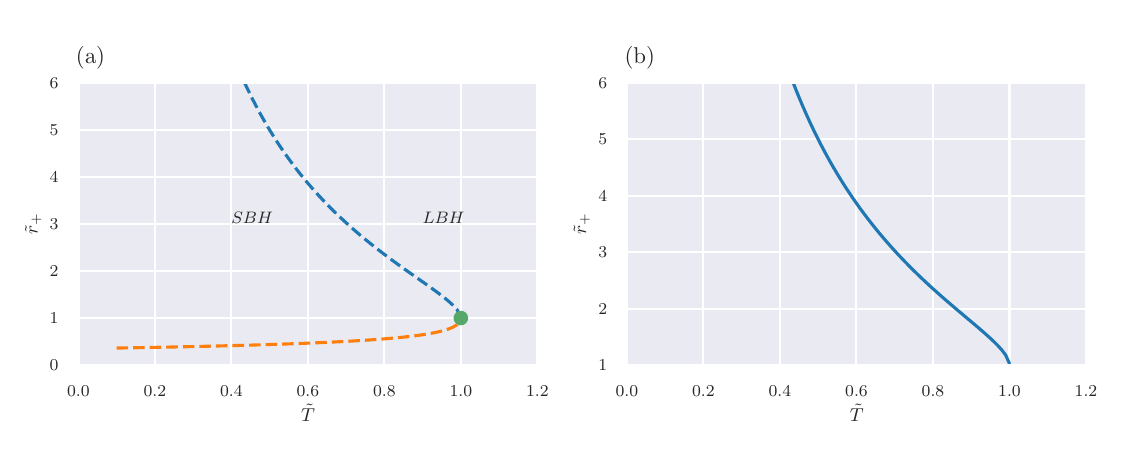}
\caption{Reduced horizon radius versus Hawking temperature. (a) An abrupt change at the critical temperature signifies a first-order phase transition. (b) The horizon radius behaves as an order parameter.}
\label{fig3}
\end{figure*}

It is well known that charged AdS black holes exhibit critical behaviour similar to the van der Waals system in conventional thermodynamics. The critical parameters associated with the black hole system are obtained by using the condition
$\partial P /\partial V = 0$ and $\partial^2 P /\partial V^2 = 0$, as,
\begin{equation}
T_c = \frac{0.02208}{q}, \quad P_c = \frac{0.0008926}{q^2}, \quad V_c = 417.157\,q^3.
\end{equation}
With the help of these, we define the reduced parameters,
\begin{equation}
\tilde{T} = \frac{T}{T_c},\quad \tilde{P} = \frac{P}{P_c},\quad \tilde{V} = \frac{V}{V_c}.
\end{equation}
In the reduced parameter space, the equation of state takes the form,
\begin{align}
\tilde{P} =& \frac{0.00978}{\bigl(85.941 \,\tilde{V}^{2/3} - 4\bigr)^2 \,\tilde{V}^{4/3}}
\Bigg[
217354\,\tilde{T}\,\sqrt{85.941\,\tilde{V}^{2/3} - 4}\,\tilde{V}^2
\nonumber\\
&- 6785.63\,\tilde{V}^{2/3} + 291582\,\tilde{V}^{4/3}
- 1.5661\times 10^6\,\tilde{V}^2 + 16\,\pi^2
\Bigg].
\label{reducedeos}
\end{align}

The reduced equation of state is independent of the NLED parameter $q$. In Fig.~\ref{fig1}(a) we plot the reduced pressure $\tilde{P}$ against reduced volume $\tilde{V}$. The isotherms exhibit oscillatory behaviour for temperatures less than the critical temperature ($T_c$), which vanishes for $T > T_c$. The same behaviour can be observed from the Gibbs free energy $\tilde{G} = M - TS$ plotted against reduced temperature $\tilde{T}$ (Fig.~\ref{fig1}(b)). It shows a swallowtail behaviour\footnote{We note that the well-known swallow-tail behavior of the $G$-$T$ curve is not properly observed in \cite{Singh:2021nvm}.}, which corresponds to the oscillatory section of the $\tilde{P}$-$\tilde{V}$ diagram. This section is unphysical, and the system follows the curve without entering the swallowtail section. The point of intersection gives the coexistence point, signifying that two phases coexist at that point. As the pressure increases, the swallowtail behaviour vanishes at the critical values. The plot of these coexistence points gives the coexistence curve, which terminates at the critical point. We fit this curve and find the fitting equation of the coexistence curve to be,
\begin{align}
\tilde{P} = &-51.1418 \,\tilde{T}^{10} + 287.367 \,\tilde{T}^9 - 701.533 \,\tilde{T}^8 + 977.194 \,\tilde{T}^7
\nonumber\\
&-857.004 \,\tilde{T}^6 + 492.65 \,\tilde{T}^5 - 187.017 \,\tilde{T}^4 + 46.3054 \,\tilde{T}^3
\nonumber\\
&-6.39543 \,\tilde{T}^2 + 0.595748 \,\tilde{T} - 0.0211245.
\end{align}
We can observe a small black hole (SBH)--large black hole (LBH) phase transition in the black hole system, with the horizon radius as the order parameter. This behaviour is similar to that of the van der Waals system of liquid--vapour phase transition.

Next, we obtain the spinodal curve, using the condition,
\begin{equation}
\bigl(\partial_{\tilde{V}} \tilde{P}\bigr)_{\tilde{T}}=0.
\end{equation}
The explicit form of the spinodal curve is,
\begin{equation}
\tilde{T}_{sp} = \frac{\mathcal{A}}{\frac{1417.65}{\sqrt[3]{\tilde{V}}\,\tilde{T}^{3/2}} 
- \frac{182752\,\sqrt[3]{\tilde{V}}}{\tilde{T}^{5/2}} },
\end{equation}
where,
\begin{align}
\mathcal{A} =& \frac{177.032}{\tilde{V}^{5/3}\,\tilde{T}^3} - \frac{44.258}{\tilde{V}^{5/3}\,\tilde{T}^2}
+ \frac{2.05992}{\tilde{V}^{7/3}\,\tilde{T}^2}
- \frac{1.75579\times 10^6 \,\sqrt[3]{\tilde{V}}}{\tilde{T}^3}
\nonumber\\
&- \frac{7607.16}{\tilde{V}\,\tilde{T}^3} + \frac{326883}{\sqrt[3]{\tilde{V}}\,\tilde{T}^3}
+ \frac{4.54747\times 10^{-13}}{\tilde{V}\,\tilde{T}^2} + \frac{10215.1}{\sqrt[3]{\tilde{V}}\,\tilde{T}^2}.
\end{align}
We plot the spinodal curve in Fig.~\ref{fig2}(a); the blue dashed lines are the spinodal curves on either side of the coexistence curve, and they terminate with it at the critical point. The area between the coexistence curve and spinodal curves signifies the metastable phases. The area below the coexistence curve and above the spinodal curve corresponds to the supercooled LBH, and the area above the coexistence curve and between the spinodal curve corresponds to the superheated SBH. Beyond the critical point, the phase transition is second order where SBH and LBH are indistinguishable, referred to as the supercritical black hole phase. The coexistence curve and spinodal curves in the $T$-$V$ plane are plotted in Fig.~\ref{fig2}(b).

As the black hole undergoes a phase transition, its radius changes abruptly, which is depicted in Fig.~\ref{fig3}(a), a characteristic feature of a first-order phase transition. The change in radius as a function of temperature is shown in Fig.~\ref{fig3}(b), indicating that the radius behaves as an order parameter. Since the horizon radius characterizes the phase transition properties of the system, one can expect a similar correlation between thermodynamics and quantities dependent on the horizon radius, such as null geodesics around the black hole, which we explore in the next section.

\section{Photon Sphere and Phase Transition}
\label{sec3}
\subsection{Geodesic equations}

\begin{figure}[b]
\centering
\includegraphics[scale=0.9]{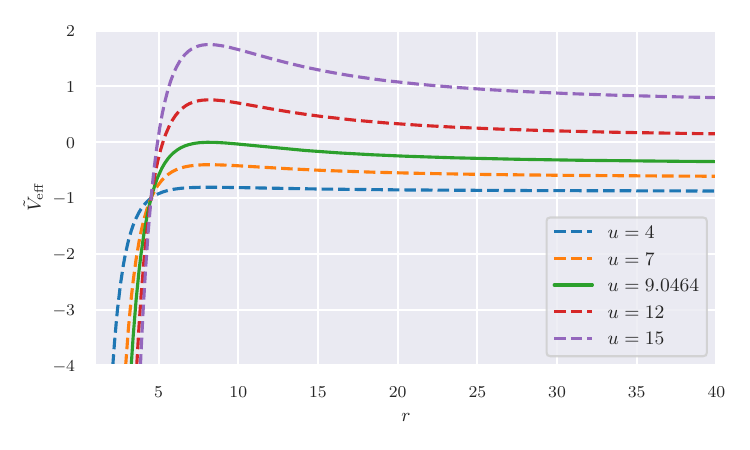}
\caption{Behaviour of effective potential of the black hole.}
\label{fig4}
\end{figure}

\begin{figure*}[t]
\centering
\includegraphics[scale=0.85]{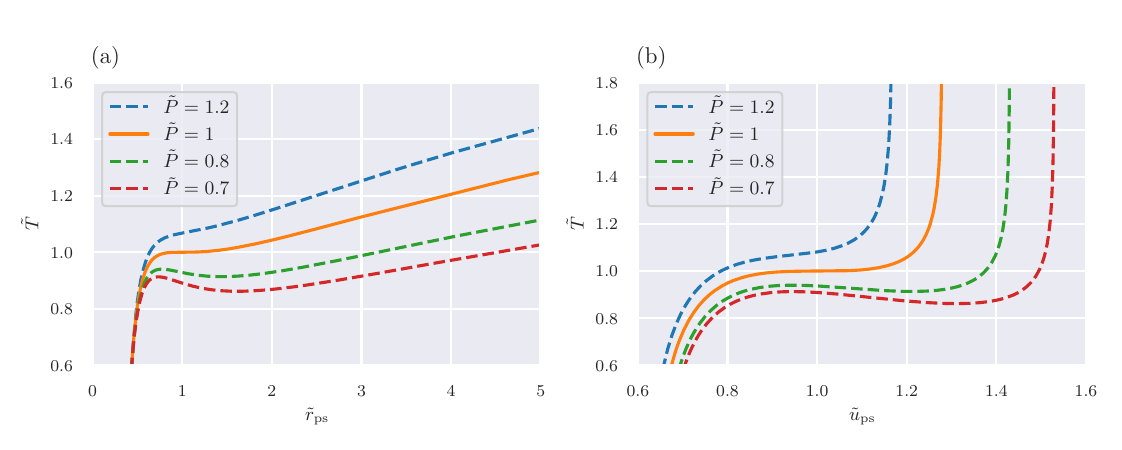}
\caption{(a) Temperature vs. photon sphere radius and (b) temperature vs. minimum impact parameter, for different values of pressure, show an oscillatory behaviour similar to vdW-like fluids.}
\label{fig5}
\end{figure*}

\begin{figure*}[t]
\centering
\includegraphics[scale=0.85]{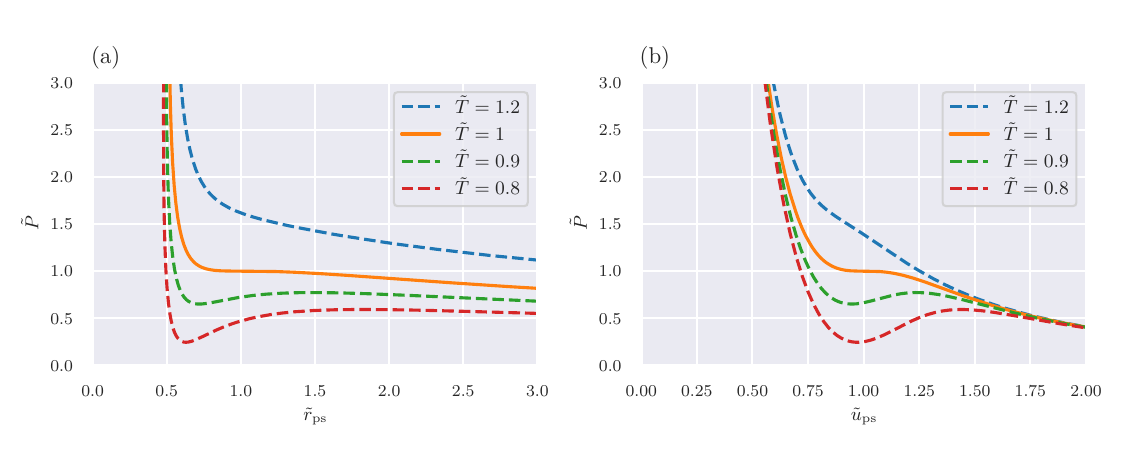}
\caption{(a) Pressure vs. photon sphere radius and (b) pressure vs. minimum impact parameter, for different values of temperature, show an oscillatory behaviour similar to vdW-like fluids.}
\label{fig6}
\end{figure*}

We now study the relationship between the phase transition and the photon sphere of the black hole \footnote{Photons in nonlinear electrodynamics propagate along null geodesics of an effective optical metric rather than of the original spacetime metric \cite{Novello:1999pg}. Nonetheless, employing the metric’s null geodesics is a widely used first-order approximation in NLED black-hole studies, because it captures the existence and qualitative evolution of unstable circular orbits making the analysis simpler \cite{Du:2022quq,Okyay:2021nnh,Uniyal:2022vdu,NaveenaKumara:2019nnt}.}. We begin by considering the equatorial ($\theta = \pi/2$) orbit of a photon around the black hole. The photon's motion is described by the following Lagrangian,
\begin{equation}
\label{lagrangian}
2 \mathcal{L} = -f(r)\dot{t}^2 + \frac{\dot{r}^2}{f(r)} + r^2 \dot{\varphi}^2,
\end{equation}
where $\dot{x}^\mu = \frac{dx^\mu}{d\lambda}$ and $\lambda$ is the affine parameter. The symmetries of spacetime lead to Killing fields $\partial_t$ and $\partial_\varphi$, which are associated with the conserved quantities of the photon’s motion: energy and orbital angular momentum. The generalized momenta of the system are given by $p_a = g_{ab} \dot{x}^b$. They are,
\begin{eqnarray}
p_t &=& -f(r)\dot{t} \equiv E,\\
p_\varphi &=& r^2 \dot{\varphi} \equiv L,\\
p_r &=& \frac{\dot{r}}{f(r)}.
\end{eqnarray}

Using these definitions, the equations of motion for $r$ and $t$ can be written as,
\begin{equation}
\dot{t} = \frac{E}{f(r)},
\end{equation}
\begin{equation}
\dot{\varphi} = \frac{L}{r^2 \sin^2\theta}.
\end{equation}
We note that the Hamiltonian of the system vanishes,
\begin{equation}
2\mathcal{H} = - E \dot{t} + L \dot{\varphi} + \frac{\dot{r}^2}{f(r)} = 0.
\end{equation}

From the above, the radial motion can be expressed as,
\begin{equation}
\dot{r}^2 + V_{eff} = 0,
\end{equation}
where the effective potential is given by, 
\begin{equation}
V_{eff} = \frac{L^2}{r^2} f(r) - E^2.
\end{equation}

The photon’s trajectory around the black hole is determined by this effective potential. The behaviour of the effective potential relative to the radius is plotted in Fig.\ref{fig4} for different values of $u$. For orbital motion to exist, $V_{eff} < 0$, as $\dot{r}^2 \geq 0$. For a particular value of angular momentum, the photon has a circular orbit forming a photon sphere. For values smaller or larger than this critical angular momentum, the photon is absorbed or scattered by the black hole, respectively. The photon orbit is characterized by,
\begin{equation}
\label{orbit}
V_{eff} = 0,\quad V'_{eff} = 0,\quad V''_{eff} < 0,
\end{equation}
where $V'_{eff} = \partial V_{eff}/\partial r$ and $V''_{eff} = \partial^2 V_{eff}/\partial r^2$. In this orbit, the radial velocity of the photon is zero. The condition $V'_{eff} = 0$ can be expanded as,
\begin{equation}
2f(r_{ps}) - r_{ps}\,\partial_r f(r_{ps}) = 0.
\label{aneqn}
\end{equation}

\begin{figure*}[t]
\centering
\includegraphics[scale=0.85]{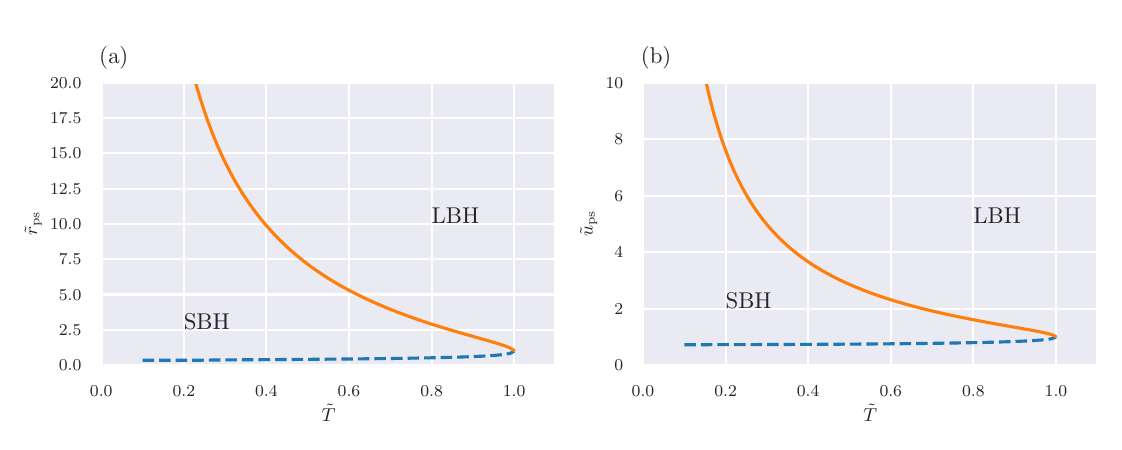}
\caption{(a) Variation of photon sphere radius with respect to temperature and (b) variation of minimum impact parameter with respect to temperature, show an abrupt change at the critical temperature, signifying a phase transition.}
\label{fig7}
\end{figure*}

\begin{figure*}[t]
\centering
\includegraphics[scale=0.85]{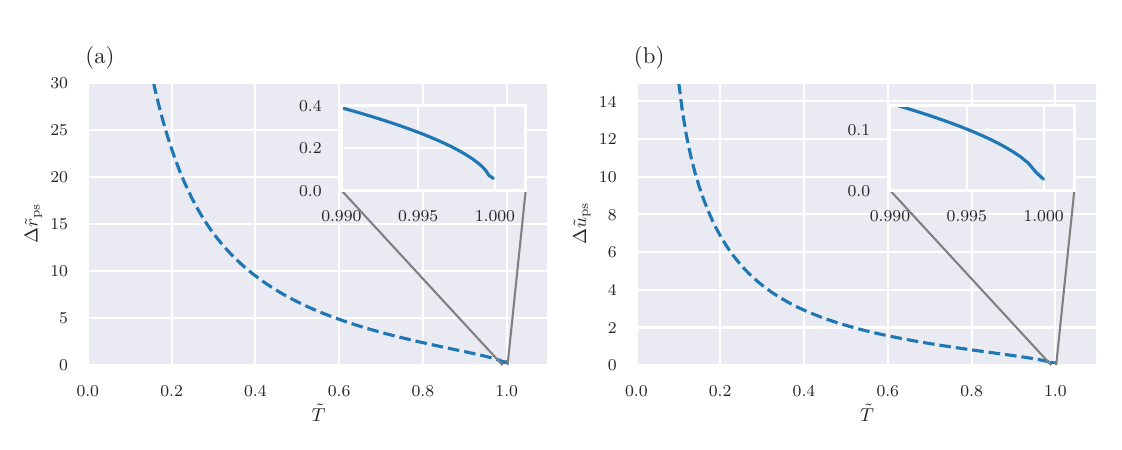}
\caption{(a) Change in photon sphere radius with respect to temperature and (b) change in the minimum impact parameter with respect to temperature, show that both the photon sphere radius and minimum impact parameter act as order parameters. Insets show the behaviour near the critical temperature.}
\label{fig8}
\end{figure*}

Substituting the metric function in this equation provides the expression for $r_{ps}$, which depends on the black hole parameters $(M, q, P)$. Another key parameter of the photon sphere is the minimum impact parameter, obtained by solving $(V_{eff} = 0)$, which reads,
\begin{equation}
u_{ps} = \frac{L_c}{E} = \left.\frac{r}{\sqrt{f(r)}} \right|_{r_{ps}}.
\label{upsequation}
\end{equation}

In a reduced parameter space, by observing the photon sphere radius and minimum impact parameter with respect to the Hawking temperature and pressure, we find the correlation between the photon sphere and black hole phase transition. The reduced impact parameter $\tilde{u}_{ps}$ is defined as, $\tilde{u}_{ps} = {u_{ps}}/{u_{psc}}$, and the reduced photon sphere radius $\tilde{r}_{ps}$ is defined as, $\tilde{r}_{ps} = {r_{ps}}/{r_{psc}}$, where ${u_{psc}}$ and ${r_{psc}}$ are the critical impact parameter and critical photon sphere radius, respectively. The isobars in $\tilde{T} - \tilde{r}_{ps}$ and $\tilde{T} - \tilde{u}_{ps}$ show a behaviour analogous to that of a van der Waals system (Fig.~\ref{fig5}). Below the critical pressure, they exhibit oscillatory behaviour; above the critical pressure, this feature disappears. Similarly, an oscillatory behaviour below the critical temperature is shown by the isotherms in the $\tilde{P} - \tilde{r}_{ps}$ and $\tilde{P} - \tilde{u}_{ps}$ planes (Fig.~\ref{fig6}). However, the increasing and decreasing trends are reversed. These behaviours are characteristic features of the van der Waals-like phase transition of the black hole, clearly illustrating the connection between photon orbit and phase transition.

\subsection{Critical behaviour of the photon sphere}
\label{secphotoncritical}

\begin{figure*}[t]
\centering
\includegraphics[width=\textwidth]{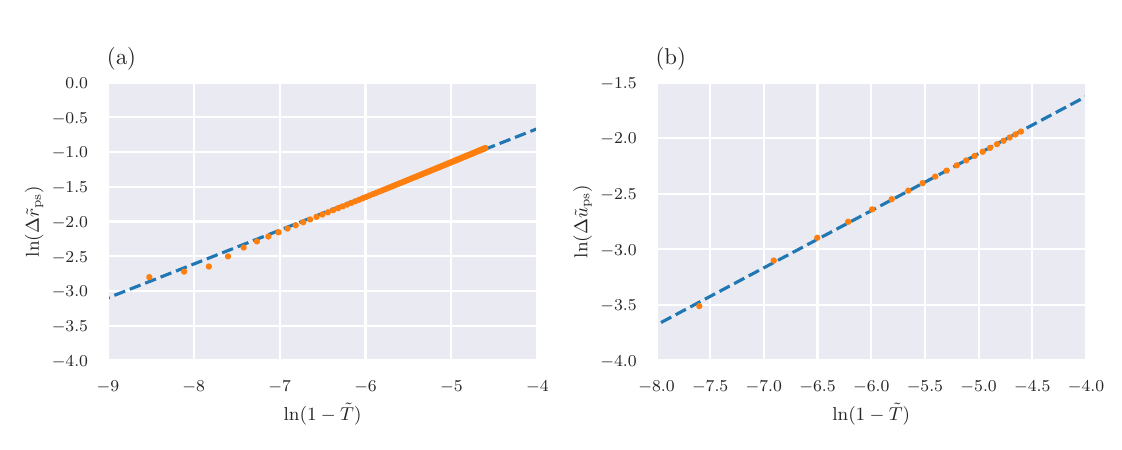}
\caption{Fitting curves for (a) radius against temperature and (b) impact parameter against the temperature of the black hole, near the critical temperature.}
\label{fig9}
\end{figure*}

The AGB black hole in AdS spacetime exhibits a van der Waals-like first-order phase transition below the critical point. At the critical point, this behaviour ceases, and a second-order phase transition is observed. In the previous subsection, we saw that the photon sphere parameters effectively capture these features, so it is reasonable to see changes in these parameters during the phase transition. We construct the equal-area law to determine the behaviour of the photon orbit radius and minimum impact parameter. This is possible because the isobars in the $\tilde{T} - \tilde{r}_{ps}$ and $\tilde{T} - \tilde{u}_{ps}$ planes resemble the isobars in the conventional $\tilde{T} - \tilde{S}$ diagrams of the black hole. The results are shown in Fig.~\ref{fig7}. Both $\tilde{r}_{ps}$ and $\tilde{u}_{ps}$ behave similarly: they increase with temperature in the coexistence SBH phase and decrease with temperature in the coexistence LBH phase. At the critical point $\tilde{T} = 1$, the values of LBH and SBH coincide.

In Fig.~\ref{fig8}, we show the change in photon orbit parameters versus the phase transition temperature. For a first-order phase transition, the change is finite, whereas at the second-order transition point, these differences vanish. The behaviour of both $\Delta \tilde{r}_{ps}$ and $\Delta \tilde{u}_{ps}$ therefore serves as an order parameter to characterize the black hole phase transition. We examine the critical behaviour associated with $\Delta \tilde{r}_{ps}$ and $\Delta \tilde{u}_{ps}$ near the critical point $\tilde{T} = 1$. Numerically, we obtain:
\begin{equation}
\Delta \tilde{r}_{ps}, \quad \Delta \tilde{u}_{ps} \;\sim\; a \times (1 - \tilde{T})^\delta.
\end{equation}
Taking the logarithm on both sides,
\begin{equation}
\ln \Delta \tilde{r}_{ps}, \quad \ln \Delta \tilde{u}_{ps} \;\sim\; \delta \ln (1 - \tilde{T}) + \ln a.
\end{equation}
Hence, $\ln \Delta \tilde{r}_{ps}$ and $\ln \Delta \tilde{u}_{ps}$ vary linearly with $\ln (1 - \tilde{T})$. By fitting the numerical data near the critical point for $0.99 \le \tilde{T} \le 0.9999$ (see Fig.~\ref{fig9}), we obtain:
\begin{equation} 
\Delta \tilde{r}_{ps} = 1.28653 \times (1-\tilde{T})^{0.487368},
\end{equation}
\begin{equation}
\Delta \tilde{u}_{ps} = 0.430121 \times (1-\tilde{T})^{0.513824}.
\end{equation}

This behaviour, i.e.\ $\Delta \tilde{r}_{ps} \sim (1-\tilde{T})^{1/2}$ and $\Delta \tilde{u}_{ps} \sim (1-\tilde{T})^{1/2}$, indicates that these quantities act as order parameters for the black hole phase transition with a critical exponent of $1/2$. The reflection of critical behaviour in $\Delta \tilde{r}_{ps}$ and $\Delta \tilde{u}_{ps}$ further demonstrates the connection between photon orbits and thermodynamic phase transitions.


\section{Ruppeiner Geometry and Interacting Microstructures}
\label{sec4}

For a spherically symmetric black hole, a novel Ruppeiner geometry method was proposed by Wei et al.~\cite{Wei2019a}, which describes the black hole’s microstructure. In this construction, the parameter space is formed by the fluctuation coordinates temperature $T$ and volume $V$. The corresponding line element is given by
\begin{equation}
dl^2 = \frac{C_V}{T^2} \, dT^2 - \frac{\left(\partial_V P\right)_T}{T} \, dV^2,
\label{line}
\end{equation}
where $C_V$ is the heat capacity at constant volume. However, the Ruppeiner curvature scalar constructed from this line element diverges, owing to the vanishing $C_V$. To address this, a normalized curvature scalar is defined as follows:
\begin{equation}
R_N = C_V \, R.
\end{equation}

\begin{figure*}[t]
\centering
\includegraphics[width=\textwidth]{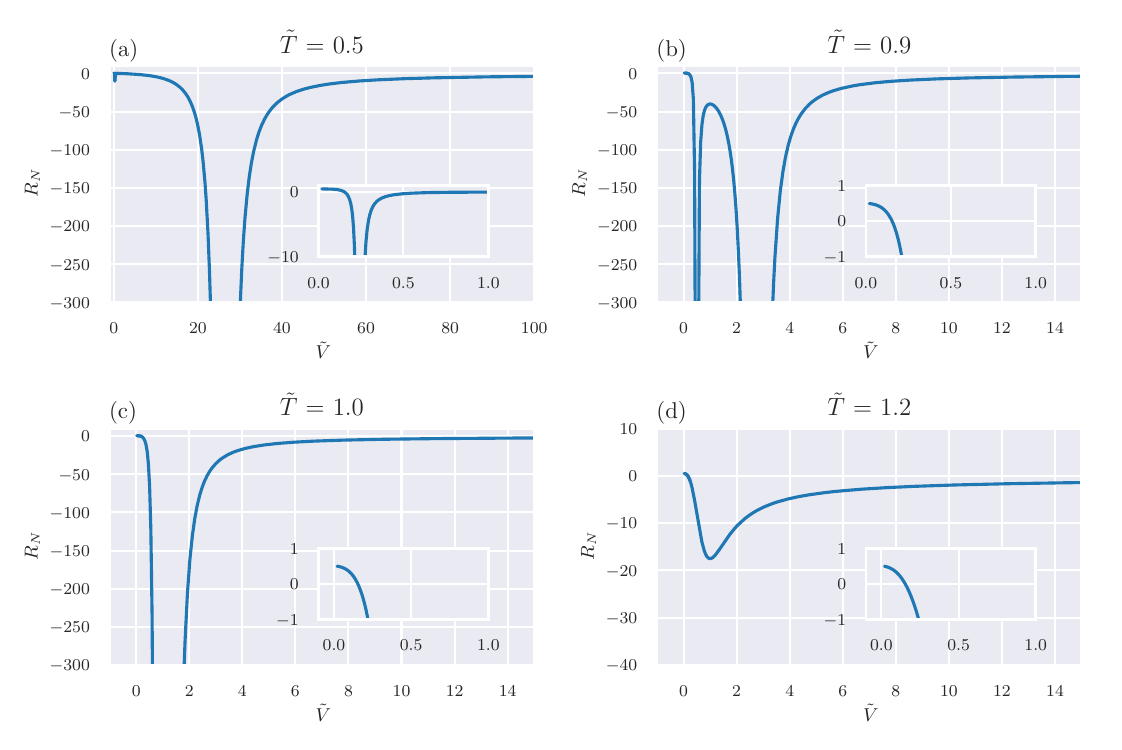}

\caption{The behaviour of the normalized curvature scalar $R_N$ against the reduced volume $\tilde{V}$ at constant temperature. The insets show the magnified view near $\tilde{V} = 0$. A repulsive interaction is observed near $\tilde{V} = 0$ in all cases. (a) $\tilde{T} = 0.5$, (b) $\tilde{T} = 0.9$, (c) $\tilde{T} = 1$, (d) $\tilde{T} = 1.2$.}
\label{fig10}
\end{figure*}

We have obtained the normalized Ruppeiner scalar $R_N$ for the ABG black hole, which is a lengthy expression. The behaviour of $R_N$ with respect to $\tilde{V}$ for different temperatures $\tilde{T}$ is shown in Fig.~\ref{fig10}. From this figure, it is clear that $R_N$ has an extremal point at $\tilde{V} = 1$. Below the critical temperature, $R_N$ shows two divergences, which merge at the critical temperature $\tilde{T} = 1$. Above that, no divergences are present. In all cases, there is a repulsive interaction at small values of $\tilde{V}$ (see the insets). However, we need to consider the thermodynamic stability of the black hole.

We obtain the sign-changing curve of $R_N$ by setting $R_N = 0$. The expression we found satisfies,
\begin{equation}
T_0 = \frac{\tilde{T}_{sp}}{2},
\end{equation}
which is a universal relation. This $T_0$ is the temperature at which $R_N$ changes sign, which is half the spinodal curve temperature.

\begin{figure*}[t]
\centering
\includegraphics[width=\textwidth]{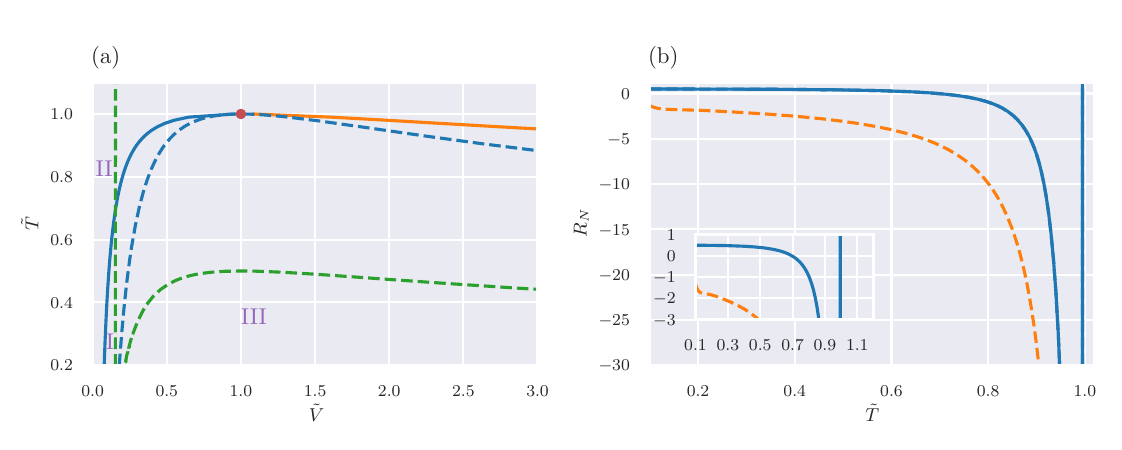}
\caption{(a) The sign-changing curve of $R_N$ (dashed green curve), coexistence curve (solid blue and orange curves), and spinodal curve (dashed blue curve). The dot indicates the critical temperature. The regions where $R_N$ is positive are highlighted as I, II, and III. (b) The normalized curvature scalar $R_N$ along the coexistence-saturated SBH and LBH phases. The blue (solid) line and the orange (dashed) line correspond to the small black hole and large black hole, respectively. The region where $R_N$ is positive is highlighted in the inset.}
\label{fig11}
\end{figure*}

In fact, along the spinodal curve, the normalized Ruppeiner scalar diverges. We note that in the region below the sign-changing curve, $R_N$ has a positive sign, indicating repulsive interaction among the black hole microstructures; this is the region III of Fig.~\ref{fig11}(a), similar to what is observed in van der Waals fluids and charged AdS black holes. However, regions lying below the spinodal curve are part of the system’s unstable states and are thus of no significance. The physically relevant regions of parameter space are above the spinodal curve, among which two regions are of interest, namely region $I$ and region $II$, as shown in Fig.~\ref{fig11}(a). As in the case of the RN-AdS black hole, region $II$ (to the left of the coexistence curve) shows repulsive interaction in the small black hole phase. The shaded region $I$, between the sign-changing and coexistence curves, is the metastable SBH phase with repulsive interaction. Elsewhere, the microstructure resembles that of a van der Waals fluid.

\begin{figure*}[t]
\centering
\includegraphics[width=\textwidth]{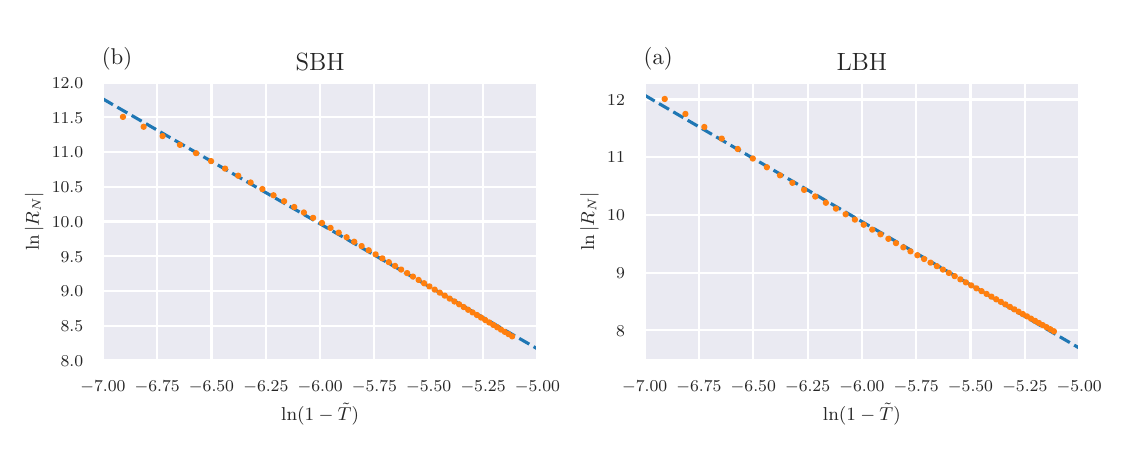}
\caption{Fitting curves for the curvature scalar of (a) the small and (b) large black hole phases.}
\label{fig12}
\end{figure*}

Next, we consider the behaviour of $R_N$ along the coexistence curve. The analytical expression for the coexistence curve is not tractable due to the complexity of the spacetime, so a numerical solution is obtained. The result is shown in Fig.~\ref{fig11}(b). As expected, the curvature scalar diverges near the critical point for both the SBH and LBH phases. At lower temperatures, there is a repulsive interaction for the SBH phase, as evidenced by the curve crossing the $R_N = 0$ line. However, at higher temperatures, the SBH phase has a dominant attractive interaction. This shows that the SBH phase behaves like an anyon gas. By contrast, the LBH phase always exhibits attractive interaction, similar to a boson gas. From these observations, we infer that at lower temperatures, during the phase transition, both the microstructure and the type of interaction change. However, at higher temperatures, the nature of the interaction remains unchanged even though the microstructure changes. In the case of a van der Waals fluid, the system always possesses a dominant attractive interaction, which remains unchanged during phase transition.

Finally, we examine the near-critical point behaviour of the curvature scalar by numerical methods. The numerical fit is obtained by assuming the following functional form:
\begin{equation}
R_N \sim (1 - \tilde{T})^p,
\end{equation}
which can be written as
\begin{equation}
\ln |R_N| = -p \,\ln(1 - \tilde{T}) + q.
\end{equation}
Performing the numerical fit for the SBH and LBH branches separately, we find
\begin{eqnarray}
\textit{SBH}:\quad \ln R_N = -1.79837 \,\ln(1 - \tilde{T}) - 0.823466, \label{sbhnum}\\
\textit{LBH}:\quad \ln R_N = -2.19002 \,\ln(1 - \tilde{T}) - 3.25666, \label{lbhnum}
\end{eqnarray}
which are shown in Fig.~\ref{fig12} along with the numerical fitting data. From the numerical study, we conclude that $p \approx 2$; we thus set $p = 2$, taking numerical errors into account. Combining these equations, we obtain
\begin{align}
R_N (1 - \tilde{T})^2 &= -\,e^{-\tfrac{0.823466 + 3.25666}{2}} \nonumber \\
&= -0.130021 \approx -\tfrac{1}{8}.
\end{align}
This agrees with the universal result for van der Waals fluids and other AdS black holes \cite{Wei2019a, Wei2019b, Wei:2019ctz, HosseiniMansoori:2020jrx}, showing that $R_N$ has a universal exponent $2$ and satisfies $R_N(1 - \tilde{T})^2 = -\tfrac{1}{8}$ near the critical point.

\section{Discussions}
\label{sec5}

In this article, we have studied the thermodynamics, the correlation between photon orbits and phase transitions, and the microstructure of the ABG black hole using Ruppeiner geometry. The black hole exhibits van der Waals-like critical behaviour, which is evident from the swallowtail behaviour of free energy. It undergoes a first-order phase transition between the small black hole (SBH) and large black hole (LBH) phases, which terminates at the critical point where the phase transition becomes second order. The coexistence curve for the system is obtained numerically from the behaviour of the Gibbs free energy.

Studying the photon sphere around the black hole establishes a correlation between the thermodynamics and gravity of the black hole. The phase transition behaviour is mirrored in the photon orbit parameters, such as the radius and minimum impact parameter. These parameters undergo a sudden change during the phase transition, and the differences $\Delta r_{ps}$ and $\Delta u_{ps}$ serve as order parameters with a critical exponent of $1/2$. Due to the complexity of the non-linearly coupled electromagnetic field, the analysis is carried out numerically. The ABG NLED parameter $q$ behaves similar to the charge parameter $q$ of the RN-AdS \cite{Wei:2017mwc}, parameter $g$ of Bardeen and Hayward black holes \cite{NaveenaKumara:2019nnt}, and to the Gauss-Bonnet coupling parameter $\alpha$ of the Einstein-Gauss-Bonnet solution \cite{Hegde:2020yrd}.

Finally, we study the underlying microstructure using the novel Ruppeiner geometry method, which shows a deviation from typical van der Waals-like systems. The curvature scalar behaviour is analysed along the coexistence curve, revealing a dominant repulsive interaction within certain intervals of the parameter space of temperature and volume. Our study shows that the large black hole phase behaves as a bosonic system, with only dominant attractive interactions, and the small black hole phase resembles an anyonic system, with both dominant repulsive and attractive interactions. The behaviour of the curvature scalar near the critical point is analysed numerically and is found to satisfy the universal relation $R_N(1 - \tilde{T})^2 = -\frac{1}{8}$.
\section*{Acknowledgments}
The authors K.H., N.K.A., and A.R.C.L. would like to thank U.G.C. Government of India for financial assistance under the UGC-NET-SRF scheme. M.S.A.'s research is supported by the ISIRD grant 9-252/2016/IITRPR/708. A.R.C.L.'s research is supported by the project grant received under the PM USHA Scheme, G.O.(Rt)No.239/2025/HEDN dated 22.02.2025.

\bibliography{BibTex.bib}

\end{document}